\documentclass[reprint,aps,pre,superscriptaddress]{revtex4-2}
\usepackage[utf8]{inputenc}
\usepackage[english]{babel}

\usepackage{bbm,bm,amsthm,amsfonts,amssymb,amsopn,amsmath,mathtools,physics,graphicx,array,multirow,url,hyperref,enumerate}
\usepackage[dvipsnames]{xcolor}
\usepackage{scalerel,color,pgf,tikz,placeins}
\usetikzlibrary{decorations.pathmorphing,decorations.markings,patterns,decorations.pathreplacing,shapes.misc}
\definecolor{colBall}{rgb}{0.27,0.45,0.77}
\definecolor{colBallEdzge}{rgb}{0,0.01,0.18}

\newcommand{\be}{\begin{equation}}
\newcommand{\ee}{\end{equation}}
\renewcommand{\vec}[1]{{\boldsymbol{\mathbf{#1}}}}
\newcommand{\ul}[1]{\underline{#1}}

\DeclareMathOperator{\Span}{Span}

\newcommand{\X}[1]{X_{#1}}
\newcommand{\Y}[1]{Y_{#1}}
\newcommand{\Z}[1]{Z_{#1}}

\newcommand{\A}[1]{A_{#1}}
\newcommand{\B}[1]{B_{#1}}
\newcommand{\ii}{\mathrm{i}}
\newcommand{\re}{\mathrm{e}}
\newcommand{\fl}{-}

\newcommand{\free}{\mathrm{(FF)}}
\newcommand{\inter}{\mathrm{(XZZ)}}
\newcommand{\pf}[1]{{\color{blue}\ifmmode\text{\footnotesize(PF) #1}\else\footnotesize{(PF) #1}\fi}}

\newcommand{\W}{\mathbb{W}}

\usepackage{pgffor}
\usepackage{xprintlen}

\begin{document}

\preprint{APS/123-QED}

\title{Stochastic strong zero modes and their dynamical manifestations}
\author{Katja Klobas} 
\affiliation{Rudolf Peierls Centre for Theoretical Physics, University of Oxford, Parks Road, Oxford, OX1 3PU, United Kingdom}

\author{Paul Fendley} 
\affiliation{Rudolf Peierls Centre for Theoretical Physics, University of Oxford, Parks Road, Oxford, OX1 3PU, United Kingdom}
\affiliation{All Souls College, University of Oxford, Oxford, OX1 4AL, United Kingdom}

\author{Juan P. Garrahan}
\affiliation{School of Physics and Astronomy, University of Nottingham, Nottingham, NG7 2RD, United Kingdom}
\affiliation{Centre for the Mathematics and Theoretical Physics of Quantum Non-equilibrium Systems, University of Nottingham, Nottingham, NG7 2RD, United Kingdom}
\affiliation{All Souls College, University of Oxford, Oxford, OX1 4AL, United Kingdom}

\date{\today} 

\begin{abstract}
     Strong zero modes (SZMs) are conserved operators localised at the edges of certain quantum spin chains, which give rise to long coherence times of edge spins. Here we define and analyse analogous operators in one-dimensional {\em classical stochastic} systems. For concreteness, we focus on chains with single occupancy and nearest-neighbour transitions, in particular particle hopping and pair creation and annihilation. For integrable choices of parameters we find the exact form of the SZM operators. Being in general non-diagonal in the classical basis, the dynamical consequences of stochastic SZMs are very different from those of their quantum counterparts. We show that the presence of a stochastic SZM is manifested through a class of exact relations between time-correlation functions, absent in the same system with periodic boundaries.
     %a large class of exact transient relations between time-correlation functions, absent in the same system with periodic boundaries.
\end{abstract}

\maketitle

Recent successes have transformed our understanding of how
long relaxation times---and potential non-ergodicity--- emerge in quantum
many-body systems (for reviews see e.g.\
\cite{Eisert2014,DAlessio2016,Nandkishore2015, Abanin2017,
Khemani2019,moudgalya2021quantum}). One simple mechanism in some systems with open
boundaries is that of a {\em strong zero mode}
(SZM)~\cite{Kitaev2001,Fendley2012,Fendley2016,Alicea2016,Kemp2017,Else2017,Vasiloiu2018,Vasiloiu2019,Vasiloiu2022,
yates2019almost,yates2020dynamics,yates2022long}.
An SZM is an operator localised at the boundary that commutes with the Hamiltonian, up to
exponentially small corrections. Its presence 
affects the structure of the whole spectrum of the Hamiltonian, resulting for
example in boundary degrees of freedom having very long coherence times~\cite{Kemp2017,Else2017,Vasiloiu2018,Vasiloiu2019,Vasiloiu2022}. 
For certain integrable spin chains, SZMs can be constructed exactly and explicitly~\cite{Fendley2016}.

In a {\em classical stochastic} system, continuous-time Markov dynamics are defined by a stochastic generator, just like a Hamiltonian generates unitary dynamics in a quantum system. While being in general non-Hermitian, stochastic generators often share many properties with Hamiltonians, thus connecting classical stochastic and quantum problems at the technical level. An example of such a connection is between the simple exclusion process and the XXZ quantum chain, see e.g.~\cite{sandow1994partially,essler1996representations,golinelli2006derivation,Gier2006}. A natural question to ask, therefore, is whether SZMs exist in classical stochastic systems, and if they do, what consequences they have for the dynamics.

Here we answer this question. For simplicity we focus on systems of particles on a one-dimensional chain with at most single occupancy per site. We consider transitions between neighbouring sites, including hopping and pair creation or annihilation. Detailed balance need not be obeyed. For certain choices of the transition rates the generators are integrable, and for these we find the explicit form of boundary localised operators that commute with the generator (either exactly or up to corrections that are exponentially small in the system size). These {\em stochastic SZMs} are non-diagonal in the classical basis, and as such do not correspond to classical observables. They represent ``hidden'' conservation laws which, as we show below, manifest themselves in the dynamics through a class of exact relations among time-correlation functions observable at finite times. 

\begin{figure}
    \includegraphics[width=0.8\columnwidth]{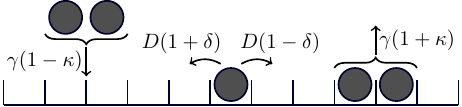}
    \caption{
        \label{fig:transitions} Schematic representation of allowed transitions.
        A particle can only hop to a neighbouring site if that site is empty. The rate to hop to the left is $D(1+\delta)$ and that to the right is $D(1-\delta)$. A pair of neighbouring particles can evaporate with rate $\gamma(1+\kappa)$, while a pair of empty sites can condense a pair of particles with rate 
        $\gamma(1-\kappa)$.
    }
\end{figure}

We study a system of particles stochastically hopping on a one-dimensional
chain of length $L$, while obeying an exclusion constraint so that each site
can be occupied by at most one particle. A particle can hop to a neighbouring
site (either left or right) if it is empty, two particles positioned on consecutive sites
can evaporate from the lattice, and two particles can condense on a pair of empty sites.
As illustrated in Fig.~\ref{fig:transitions}, the left- and right-hopping transitions have rates $D(1+\delta)$ and $D(1-\delta)$ respectively, while evaporation and condensation
occur with rates $\gamma(1+\kappa)$, and $\gamma(1-\kappa)$.  At the edges we typically assume open boundary conditions, where the first and last site each have only one nearest neighbour \cite{Note3}. 
\footnotetext[3]{
  All transitions in the models we consider involve two neighbouring sites. This means that in the case of open boudaries, the leftmost set of transitions (hopping/pair-condensation/pair-annihilation) are between sites 1 and 2, while the rightmost between sites $L-1$ and $L$. The local generator (that is, the one for transitions between sites $i$ and $i+1$) is stochastic in itself. This means that we could even have chosen site dependent rates (see Fig.~\ref{fig:transitions}) while keeping the overall generator stochastic. See Eqs.~\eqref{eq:defWff} and~\eqref{eq:defWXZZ}.
}

At each time the configuration of the system can be expressed in
terms of a $L$-tuple $\ul{n}=(n_1,n_2,\ldots,n_{L})\in\mathbb{Z}_2^L$, where
$n_j=1$ if there is a particle on site $j$ and $n_j=0$ when empty. To describe
dynamics of \emph{macroscopic} states (i.e.\ probability distributions) we use
bra-ket notation, $\ket{p}=[p_0,p_1,\ldots,p_{2^L-1}]\in\mathbb{R}^{2^L}$,
where each component $p_n\ge 0$ represents a probability of the configuration
given by the binary representation of the subscript $n$, and the sum of all
components is one, $\sum_{n} p_n=1$. Diagonal operators represent \emph{observables},
i.e.\ quantities that can be measured. Their expectation values are by definition
given by the sum $\expval{a}_{p}=\sum_{n} a_{n,n} p_{n}=\mel{\fl}{a}{p}$, where
we introduced the \emph{flat state} $\bra{\fl}=[1\ 1]^{\otimes L}$. The
normalization condition for $\ket{p}$ then can be equivalently expressed as $\braket{\fl}{p}=1$. 
Encoding the stochastic transitions with
a generator $\W$ means that an initial state $\ket{p}$ evolves in time as $\ket{p(t)}=\mathrm{e}^{t \W}\ket{p}$. 
Conservation of probability under time evolution requires
$\bra{\fl}\mathrm{e}^{t \W}=\bra{\fl}$. This notation gives convenient expressions for
more complicated objects, such as the expectation value at time $t$ after starting from some
non-stationary initial state $\mel{\fl}{a\mathrm{e}^{t \W}}{p}$, or correlation functions
between multiple observables at different times,
$\mel{\fl}{b\,\mathrm{e}^{(t_2-t_{1}) \W}a\,\mathrm{e}^{t_1 \W}}{p}$.

We restrict the discussion to two different \emph{integrable} limits of the
generator $\W$ (see e.g.~\cite{Stinchcombe2001}), for which the Hamiltonian
counterparts are known to exhibit conserved
edge modes~\cite{Fendley2012,Fendley2016,Note1}:

\footnotetext[1]{Note that these two are not necessarily the only stochastic
models that exhibit an almost conserved
edge mode~\cite{Kemp2017,Else2017}. However, these are the only
two regimes for which the explicit closed-form expression of the edge modes is
known, and we rely on that.}

\vspace{.2cm}

%\begin{enumerate}[(i)]
%    \item 
    {\it (i)}\quad In the regime $\gamma=D$ the generator is quadratic in
        fermionic operators (see Sec.~\ref{sec:FFdiag}), so we refer to
        it as the \emph{free-fermionic model}. The stochastic generator with open boundaries
        has the form  
        \be \label{eq:defWff}
        \begin{aligned}
            \mkern-20mu
            \W^{\free}=
            \sum_{j=1}^{L-1}
           & \Big[\X{j}\X{j+1}+\kappa \Z{j}
            +i\frac{\kappa+\delta}{2}\X{j}\Y{j+1}\mkern-20mu
            \\\mkern-20mu
            &+i\frac{\kappa-\delta}{2}\Y{j}\X{j+1}-1
            \Big]+\frac{\kappa+\delta}{2}\left(\Z{L}-\Z{1}\right)%-(L-1),
            \mkern-20mu
        \end{aligned}
        \ee
        where $\X{j}$, $\Y{j}$, and $\Z{j}$ denote Pauli matrices acting on the
        site $j$.  Without loss of generality, we rescaled the unit of time so that $D=1$.  Note that when
        $\delta \neq 0$, hopping is asymmetric and $\W$ does not obey detailed
        balance.
   % \item 
   
   \vspace{0.2cm}
   
      {\it (ii)}\quad
      The second integrable regime arises when $\kappa=\delta=0$, i.e.\
        there is no asymmetry between the left and right hopping, and the rates
        for condensation and evaporation are the same. This model was
        studied with periodic boundaries in
        Ref.~\cite{Grynberg1994}, and, more recently, solutions to the
        boundary-driven setup have been found~\cite{Crampe2014}. In
        this case the generator takes a form of a rotated anisotropic
        Heisenberg XXZ Hamiltonian,
        \be\label{eq:defWXZZ}
        \begin{aligned}
            \W^{\inter}=
            \sum_{j=1}^{L-1}&\Big[
        \frac{1-\gamma}{2}\left(\Y{j}\Y{j+1}+\Z{j}\Z{j+1}\right)\\
        &+\frac{1+\gamma}{2}\big(\X{j}\X{j+1}-1\big)\Big] \ ,
      %  -(L-1)\frac{1+\gamma}{2},
        \end{aligned}
        \ee
       so we refer to it as XZZ model. We again chose $D=1$.
%\end{enumerate}

\vspace{.2cm}

In analogy to the quantum setting, a \emph{conserved edge mode} $\Psi$ is an
operator that commutes with the stochastic generator, $[\Psi,\W]=0$, squares into identity,
$\Psi^2=1$, and is localised at an edge---its local densities that
involve sites far from the edge are exponentially suppressed.

In the 
case of $\W^{\free}$ we take advantage of the free-fermionic form to straightforwardly
find the expression for $\Psi^{\free}$ (see Sec.~\ref{sec:ZeroModes} for the derivation),
\be \label{eq:PsiExpr}
\Psi^{\free} = \sum_{j=1}^L \lambda^{j-1} \mu_{j-1}\left(\X{j} + i \lambda \Y{j}\right),
\ee
where the disorder operator $\mu_{j}=\prod_{k=1}^{j} \Z{k}$ is a string of $\Z{k}$ originating at the left edge, and the parameter $\lambda$ is expressed in
terms of $\kappa$ and $\delta$ as
\be \label{eq:lambdaDef}
\lambda=\frac{1-\sqrt{1+\delta^2-\kappa^2}}{\delta+\kappa} 
\ee
with $\abs{\lambda}\le 1$.
This edge mode is \emph{exactly} conserved, i.e.\ $[\W^{\free},\,\Psi^{\free} ]=0$ with no corrections. For simplicity we neglect
exponentially small corrections to the normalization:  $\left.\Psi^{\free}\right.^2=1+\mathcal{O}(\lambda^L)$.

The XZZ generator~\eqref{eq:defWXZZ} is Hermitian and has exactly the same form as the
XYZ Hamiltonian with appropriately chosen couplings, therefore we can directly adapt the
exact form of Ref.~\cite{Fendley2016} to obtain
\be
\begin{aligned} \label{eq:XYZedgemode}
    &\Psi^{\inter}=\sum_{S=0}^{\infty}
    \ \smashoperator[r]{\sum_{1\le a_1< \ldots < a_{2S}<b\le L}}
    \mkern12mu
    \lambda^{2(b-1)}\left(1-\lambda^2\right)\left(1-\frac{1}{\lambda^2}\right)^S\\
    &\times \lambda^{-\sum_{j=1}^{2S} (-1)^j a_{j}}
    \X{b}\prod_{j=1}^{S}
    \left(
    \Y{a_{2j-1}}\Y{a_{2j}}
    +\Z{a_{2j-1}}\Z{a_{2j}}
    \right),
\end{aligned}
\ee
where the value of $\lambda$, $\abs{\lambda}\le 1$, is now given by
\be
\lambda=\frac{1-\gamma}{1+\gamma}.
\ee
Unlike the free-fermion case, the edge mode now no longer exactly commutes with
the stochastic generator, but rather does so up to corrections of the order
$\mathcal{O}(\lambda^L)$. %, and the same holds for the normalization.
%It is evident that i
In both cases \eqref{eq:PsiExpr} and \eqref{eq:XYZedgemode}, $\abs{\lambda}\le 1$ implies the exponential suppression of local densities on sites far from the edge of the lattice, making the SZM localised at the boundary.

Since neither SZM is diagonal, they cannot
be directly interpreted as classical observables. Their effect on the dynamics therefore is not immediately obvious.  A key observation is that the expectation value of an off-diagonal operator $A$
always can be interpreted as an expectation value of a corresponding \emph{diagonal}
operator $\hat{A}$ defined by 
\be \label{eq:defCorrDiag}
\langle{\fl}|{A}=\langle{\fl}|{\hat{A}}\quad \implies\quad \mel{\fl}{A}{p(t)}=\mel{\fl}{\hat{A}}{p(t)}\ .
\ee
%so that 
%\be
%\mel{\fl}{A}{p(t)}=\mel{\fl}{\hat{A}}{p(t)}\ .
%\ee
Pauli operators obey the two simple identities
\be \label{eq:pauliIdentities}
\begin{bmatrix} 1 & 1 \end{bmatrix} \X{j} = \begin{bmatrix} 1 & 1 \end{bmatrix}
    \quad \text{and}\quad
\begin{bmatrix} 1 & 1 \end{bmatrix} \Y{j} = \ii \begin{bmatrix} 1 & 1 \end{bmatrix}\Z{j}\,,
\ee
which can be linearly extended to provide the diagonal operator
$\hat{A}$ for an arbitrary $A$. Therefore, the existence of a non-diagonal operator $\Psi$ 
commuting with $\W$ implies the existence of a \emph{classical observable} $\hat{\Psi}$
whose expectation value does not change with time,
\be
\begin{aligned}
    \mel{\fl}{\hat{\Psi}\mathrm{e}^{t\W}}{p}&=
\mel{\fl}{\Psi\mathrm{e}^{t\W}}{p}=
    \mel{\fl}{\mathrm{e}^{t \W}\Psi}{p}=
    \mel{\fl}{\Psi}{p}\\&=
\mel{\fl}{\hat{\Psi}}{p},
\end{aligned}
\ee
where we utilised the defining property~\eqref{eq:defCorrDiag} and
the conservation of probability. 

In our cases, a little more work is needed. Indeed, the identities~\eqref{eq:pauliIdentities} imply that $\bra{\fl}$
is (up to terms exponentially small in $L$) a left eigenvector of both
$\Psi^{\free}$, and $\Psi^{\inter}$,
\be\label{eq:flatPsi}
\bra{\fl}\Psi=\bra{\fl},
\ee
and therefore the conservation of $\mel{\fl}{\hat{\Psi}}{p(t)}$ gives us no
meaningful restriction on the dynamics.

Nonetheless, it is possible to define a \emph{dynamical protocol}, under which
the existence of the boundary mode gives nontrivial effects. We require
the initial state $\ket{\alpha}$ to be an eigenvector of $\Psi$
with eigenvalue $1$:
$\Psi\ket{\alpha}=\ket{\alpha}$ \cite{Note2}. 
The conservation of $\Psi$ implies the existence of observables whose expectation value remains constant after starting from these states.
%To demonstrate this phenomenon, we start with \
A general expectation value of an
observable $a$ at time $t$ obeys
\be\label{eq:genQuenchRel}
\begin{aligned}
    \mkern-8mu\mel{\fl}{a\,\re^{t \W}}{\alpha} \mkern-4mu &=
    \mkern-4mu\mel{\fl}{a\,\re^{t \W}\Psi^2}{\alpha}=
    \mkern-4mu\mel{\fl}{a\,\Psi\re^{t \W}}{\alpha}\mkern-4mu\\
    &=\mkern-2mu-\mkern-2mu\mel{\fl}{\Psi a\,\re^{t \W}}{\alpha}
    \mkern-4mu+\mkern-6mu\mel{\fl}{\{a,\Psi\}\re^{t \W}}{\alpha},\mkern-13mu
\end{aligned}
\ee
which follows from the normalization $\Psi^2=1$, the definition of $\ket{\alpha}$, and the conservation of the edge mode. Because $\bra{\fl}$ is the left eigenvector of $\Psi$ (cf.\
Eq.~\eqref{eq:flatPsi}), we obtain a connection between the expectation value
of $a$ at any time $t$ and that of its anticommutator %with the edge mode
$\{a,\Psi\}$:%, namely
\be \label{eq:genQuenchAnticomm}
    \mel{\fl}{a\,\re^{t \W}}{\alpha} =
    \tfrac{1}{2}\mel{\fl}{\{a,\Psi\}\re^{t \W}}{\alpha}.
\ee
This general identity can now be used to obtain some nontrivial constraints on dynamics.

\footnotetext[2]{We remark that $\Psi^2=1$ implies that the spectrum of $\Psi$
consists only of (highly degenerate) eigenvalues $1$ and $-1$.}

Let us start with the free-fermionic model, and consider $a=\Z{j}$. After a series of
elementary manipulations similar to the ones of Eq.~\eqref{eq:flatPsi}, one obtains
\[
\frac{1}{2}\bra{\fl}\big\{\Z{j},\Psi^{\free}\big\}=\bra{\fl}\Z{j}
-\lambda^{j-1}\Big(\bra{\fl}\mu_j-\lambda\bra{\fl}\mu_{j-1}\Big),
\]
which together with~\eqref{eq:genQuenchAnticomm} implies
\be \label{eq:FFrelation}
\begin{aligned}
    \mel{\fl}{\Z{1}\re^{t \W^{\free}}}{\alpha}&=
    \lambda \mel{\fl}{\re^{t \W^{\free}}}{\alpha}=\lambda,\\
    \mel{\fl}{\mu_{k}\re^{t \W^{\free}}}{\alpha}&=
    \lambda \mel{\fl}{\mu_{k-1}\re^{t \W^{\free}}}{\alpha}=\lambda^{k}.
\end{aligned}
\ee
The second equality in both rows follows from the conservation of probabilities,
$\bra{\fl}\re^{t \W}=\bra{\fl}$ and the normalization of the initial state, $\braket{\fl}{\alpha}=1$.
The expectation values of~$\mu_{k}=\prod_{j=1}^{k} \Z{j}$ are therefore constant in time, even though the initial state is \emph{not} stationary and the
system must undergo nontrivial dynamics before relaxing.
For $t=0$ the
relation~\eqref{eq:FFrelation} is the property of the initial state and does not depend
on whether or not~$\Psi^{\free}$ is conserved: the surprising consequence of the existence
of the edge mode is that it holds also when $t>0$.

The XZZ regime can be treated analogously, with only the precise relations changing
due to the different form of the edge mode. The left-action of the anticommutator on the
flat state obeys
\be \label{eq:interAnticomm}
\begin{aligned}
\tfrac{1}{2}
    \bra{\fl}\{\Z{j},\Psi^{\inter}\}&=
    %\bra
    \langle{\fl}|
    \Z{j}\,-\,%(1-\lambda^2)\lambda^{j-2}  
    \big(\lambda^{j-2}-\lambda^j\big)  \langle{\fl}|\chi
\end{aligned}
\ee
%where we introduced~$\chi$ to denote a sum of $\Z{j}$ with exponentially decaying coefficients,
where $\chi$ is a sum over the $\Z{j}$ with coefficients decaying exponentially away from the edge:
\be \label{eq:defChi}
\chi = \sum_{j=1}^L \lambda^j \Z{j}.
\ee
Inserting~\eqref{eq:interAnticomm} into~\eqref{eq:genQuenchAnticomm} immediately gives us 
the dynamical restriction for the XZZ case: the expectation value of~$\chi$ is forced to
be zero at all times, i.e.\
\be \label{eq:interrelation}
\mel{\fl}{\chi\re^{t \W}}{\alpha} = 0.
\ee

Equations~\eqref{eq:FFrelation} and~\eqref{eq:interrelation} provide nontrivial dynamical
constraints holding in the presence of the edge mode. %  for finite times $t>0$.
A few remarks are in order.  First, %all the equalities here are assumed to hold up to corrections that are 
corrections exponentially small in the system size have been ignored. Therefore one might expect that these constraints only hold up to times
of the order of magnitude $1/\lambda^{L}$. However, one can show
(see Sec.~\ref{sec:stationary}) that these expectation values coincide with the values in the
stationary state, which implies the broader applicability of the constraints.
Second, this dynamical protocol only makes sense if we can find appropriate
eigenvectors~$\ket{\alpha}$ that can be interpreted as valid probability
distributions. Since they need to satisfy the non-negativity condition, their
existence is not a priori obvious. While we have not been able to characterise
the full set of valid initial states, we have found several representative
examples (see the discussion in Sec.~\ref{sec:eigenvectors}) that we use in the numerical
demonstrations below.

\begin{figure}[h!]
    \includegraphics[width=\columnwidth]{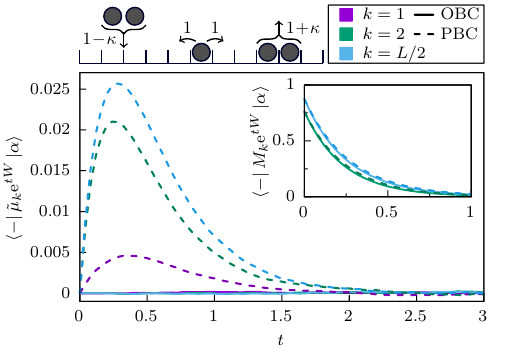}
    \caption{\label{fig:fig2}
    Dynamics of disorder operators
    $\tilde{\mu}_k=\mu_k -\lambda^k =\Z{1}\Z{2}\cdots \Z{k}-\lambda^k$ in
    the \emph{free-fermionic} model. The initial state $\ket{\alpha}$ is given
    in Eq.~\eqref{eq:freeIS}, while $\lambda$ is the expectation value of
    $\Z{k}$ in the stationary state. For open boundary conditions,
    the expectation values are restricted as in\ Eq.~\eqref{eq:FFrelation} due to the existence of the boundary
    mode, so that there is no evolution in such quantities. In contrast,
     the expectation values are unconstrained in the periodic case, and
    they undergo nontrivial time evolution. The time dependence of generic
    observables is not constrained, and they show qualitatively similar
    behaviour in both cases, as is shown in the inset for the rescaled
    magnetization $M_k=\Z{k}-\lambda$. In this example we consider symmetric
    hopping ($\delta=0$), the asymmetry between pair-annihilation and creation
    rates is $\kappa=0.25$, the system size is $L=20$, and the number of
    Monte-Carlo trajectories is $10^9$.     }
\end{figure}

To demonstrate explicitly that relations~\eqref{eq:FFrelation} and~\eqref{eq:interrelation}
represent nontrivial constraints on the time evolution, we simulate this dynamical protocol
using Monte Carlo sampling of trajectories. For clarity, we restrict the discussion to
the case of \emph{symmetric hopping} --- i.e.\ we assume $\delta=0$ in \emph{both}
regimes. The stationary state is then the same for
both periodic and open boundary conditions, while the edge mode is only conserved in
the latter case. Changing boundary conditions therefore 
gives a direct probe of the validity of the dynamical constraints arising from the
edge mode. 
The initial state in the free fermionic case is% given by
\be \label{eq:freeIS}
    \ket*{\alpha^{\free}}=\frac{1+\Psi^{\free}}{2} \begin{bmatrix}1\\0\end{bmatrix}^{\otimes L},
\ee
while the \emph{interacting} initial state is chosen as
\be \label{eq:intIS}
    \ket*{\alpha^{\inter}}=\frac{1+\Psi^{\inter}}{2} 
    \begin{bmatrix}\frac{1}{4}\\[0.5em]\frac{3}{4}\end{bmatrix}\otimes 
        \begin{bmatrix}\frac{1}{2}\\[0.5em]\frac{1}{2}\end{bmatrix}^{\otimes L-1}.
\ee
We note that these are just two concrete choices, and the full family of possible 
initial states is very large due to the high degeneracy of the spectra of $\Psi^{\free}$, 
and $\Psi^{\inter}$. See Sec.~\ref{sec:eigenvectors} for more details.

\begin{figure}[th]
    \includegraphics[width=\columnwidth]{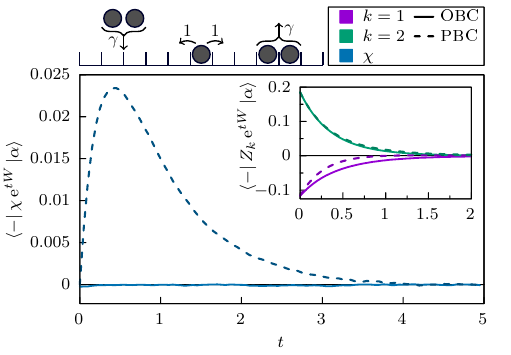}
    \caption{\label{fig:fig3}
      Expectation values of $\chi$ (defined in Eq.~\eqref{eq:defChi}) in the
      \emph{interacting} regime of the model.  When the boundary conditions are
      open, the expectation value is constrained by the existence of the edge
      SZM (Eq.~\eqref{eq:interrelation}), while the system with periodic
      boundaries exhibits nontrivial dynamics. For comparison, the dynamics of
      local magnetization $\Z{k}$ in the inset show no qualitative difference
      between the two boundary conditions. The initial state $\ket{\alpha}$ is
      given in Eq.~\eqref{eq:intIS}, the annihilation/creation rate is
      $\gamma=0.35$, the system size is $L=20$, and the number of Monte-Carlo
      trajectories is $10^8$.
   }
\end{figure}

\begin{figure}[th]
    \includegraphics[width=\columnwidth]{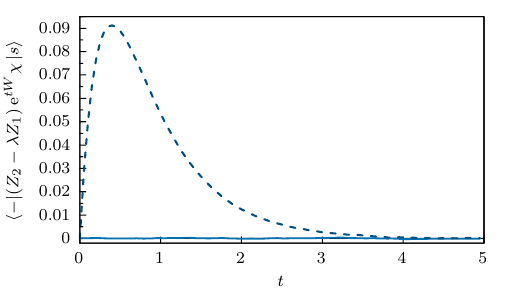}
    \caption{\label{fig:fig4} 
        Dynamical correlation function between $\chi$ and
        $\lambda^k\Z{j}-\lambda^j \Z{j}$ in the stationary state
        $\ket{s}=2^{-L}\ket{\fl}$ for the interacting regime of the model 
        (with same numerical parameters as in Fig.~\ref{fig:fig3}). The vanishing at all times for the open boundaries case illustrates the identity~\eqref{D3} that follows from the stochastic zero mode. 
   }
\end{figure}

The behaviour in the {free-fermionic} regime is shown in Fig.~\ref{fig:fig2}, where
we compare the dynamics of the expectation value~\eqref{eq:FFrelation} between
open and periodic boundaries. In both cases the initial
value is equal to the stationary value, but
the state itself is \emph{not} stationary. Therefore for periodic boundaries the
expectation value shows nontrivial dynamics, while in the open case the
edge mode prevents it from changing. The dynamics of
quantities not restricted by Eq.~\eqref{eq:FFrelation}) does not strongly
depend on the boundary conditions, as demonstrated in the inset,
where we compare the expectation value of $\Z{j}$ at two sites: one close to the edge
and one in the bulk.

Analogous behaviour can be observed in the interacting XZZ regime in
Fig.~\ref{fig:fig3}. The existence of conserved $\Psi^{\inter}$ forces the
expectation value of $\chi$ to stay at zero, see Eq.~\eqref{eq:interrelation},
while in the case of periodic boundaries there is no such restriction and
$\chi$ exhibits nontrivial dynamics.  However, as shown in the inset, the
dynamics of generic observables shows no qualitative difference between the
different boundary conditions.

In this paper we have generalised the concept of strong zero modes from quantum spin
chains to one-dimensional classical stochastic systems. For choices of
parameters that make the stochastic generators integrable we were able to
obtain the SZMs exactly. In contrast to the quantum case, the conservation of a
stochastic SZM cannot be observed directly in the dynamics, manifesting instead
as specific constraints in time correlation functions. As far as we are aware
these \emph{hidden} conservation laws in systems with open boundaries were not
identified before.

Relations~\eqref{eq:FFrelation} and~\eqref{eq:interrelation}
are just two examples of a large number of dynamical relation following from the existence of
edge modes. For example, for the case of $W^{\inter}$, a similar mechanism restricts the
dynamics of a wide class of dynamical correlation functions in the stationary state.
In particular, as we shown in Sec.~\ref{sec:correlationME}, the equilibrium time-correlation functions 
\be
\label{D3}
\mel{\fl}{\{A,\Psi\}e^{t W} B}{\fl}=\mel{\fl}{A e^{t W}\{B,\Psi\}}{\fl}
\ee
are identical --- up to times of the order of magnitude $1/\lambda^L$ --- for
any two observables $A$ and $B$.  In Fig.~\ref{fig:fig3} we plot the specific
case of $A=\Z{1}$ and $B=\Z{2}$, where \eqref{D3} reduces to
$\mel{\fl}{(\Z{2} - \lambda \Z{1}) e^{tW^{\inter}}\chi}{\fl} = 0$. 

Many questions remain. One is on the fate of SZMs away from integrability. Our results explicitly depend on the precise form of the SZMs, but
typically the physics of these models shows no
qualitative change when the stochastic rates are tuned to the integrable point.
A related question is whether for non-integrable stochastic spin chains, e.g.\ those in Ref.~\cite{Tailleur2008}, SZMs are
only conserved parametrically, as occurs in non-integrable quantum
systems~\cite{Kemp2017,Else2017}, and if so, how these ``almost'' SZMs manifest themselves in the dynamics.
A more general issue is to describe the dynamical consequences of other conserved
non-diagonal operators in classical stochastic models. For instance, setting the
condensation and evaporation rates to $0$, our model reduces to the asymmetric
simple exclusion process~\cite{spitzer1970interaction,derrida1998exactly,Blythe2007},
which can be mapped to the XXZ Heisenberg Hamiltonian by a similarity transformation.
This mapping implies the existence of an infinite number of non-diagonal local conserved 
operators that are obtained from the corresponding transfer matrix ~\cite{grabowski1995structure,faddeev1996algebraic,ilievski2016quasilocal}.
It would be very interesting to understand how they constrain the stochastic classical
dynamics.

\bigskip
\acknowledgments
This work has been supported by the EPSRC Grant no.\ EP/S020527/1
(KK, PF), EPSRC Grant no.\ EP/R04421X/1 (JPG) and the Leverhulme
Trust Grant No.\ RPG-2018-181 (JPG).

\bibliography{bibliography}

\onecolumngrid
\appendix

\bigskip

\bigskip

\begin{center}
    {\bf SUPPLEMENTAL MATERIAL}
\end{center}

\section{Diagonal form of the free-fermionic generator}\label{sec:FFdiag}
To diagonalize the generator~$\W^{\free}$ (cf.~\eqref{eq:defWff}), we first
observe that it can be put in a form that is quadratic in fermionic operators,
\be \label{eq:fermionicGenerator}
    \W^{\free}=\sum_{j=1}^{L-1}\Big(
    -\ii \kappa \A{j} \B{j}
    -\ii \B{j} \A{j+1}
    +\frac{\delta+\kappa}{2} \B{j} \B{j+1}
    +\frac{\delta-\kappa}{2} \A{j} \A{j+1}
\Big)
    +\ii \frac{\delta+\kappa}{2}\left(\A{1} \B{1}-\A{L}\B{L}\right)
    -(L-1),
\ee
where we introduced Majorana operators $\A{j}$ and $\B{j}$ as
\be
    \A{j}=\Z{1} \Z{2} \cdots \Z{j-1}\,\X{j},\qquad
    \B{j}=\Z{1} \Z{2} \cdots \Z{j-1}\,\Y{j},\qquad
    \{\A{j},\B{k}\}=0,\qquad 
    \{\A{j},\A{k}\}=\{\B{j},\B{k}\}=2 \delta_{j,k}.
\ee
The simplest way to find the spectrum of $\W^{\free}$ is to find operators $\phi$ that
obey
\be \label{eq:commRelDiag}
\left[\W^{\free},\phi\right]=2 \varepsilon \phi.
\ee
These operators are constructed as linear combinations of Majorana operators,
\be
\phi(\underbrace{\alpha_1,\beta_1,\alpha_2,\ldots,\beta_L}_{\vec{\mu}})=
\sum_{j}\left(\alpha_{j}\A{j}+\beta_j\B{j}\right),
\ee
since commuting a bilinear in fermions with a linear combination always yields
another linear combination. A convenient way of presenting $[\W^{\free},\phi]$
with such a $\phi=\phi(\vec{\mu})$ is to introduce the corresponding matrix $M_{L}$
acting on the $2L$-dimensional vector space such that
\be \label{eq:defML}
\phi(\vec{\mu}^{\prime})=[\W^{\free},\phi(\vec{\mu})]
\ \longleftrightarrow\ 
\vec{\mu}^{\prime}=2 M_L \vec{\mu}.
\ee
The possible values of $\varepsilon$ are the eigenvalues of $M_L$. In the Majorana basis
the matrix is skew-symmetric, therefore its eigenvalues come in pairs,
$(\varepsilon,-\varepsilon)$, and we label them as
\be
\varepsilon_{-k}=-\varepsilon_k,\qquad 
\Re(\varepsilon_{k})\ge 0,\qquad k=1,2,\ldots,L,
\ee
while the corresponding eigenvectors are labelled as $\vec{\mu}_k$. The matrix
has no other obvious special structure, so the eigenvalues are not guaranteed to be real,
and, indeed, in general they have a non-trivial imaginary part. Furthermore, we note
that if $\vec{\mu}_k$ is a \emph{right} eigenvector of $M_L$ corresponding to eigenvalue
$\varepsilon_k$, then it is also the \emph{left} eigenvector corresponding to eigenvalue
$\varepsilon_{-k}=-\varepsilon_k$, which follows directly from the skew-symmetric structure
of $M_L$,
\be
\vec{\mu}_k M_L=-\vec{\mu}_k M_L^{T}=
-\left(M_L\vec{\mu}_k\right)^{T}=-\varepsilon_k\vec{\mu}_k.
\ee
Therefore, the eigenbasis can be chosen so that the following holds,
\be \label{eq:orthogonalityLeftRight}
\vec{\mu}_k\cdot\vec{\mu}_q=\delta_{k,-q}.
\ee
Defining now $\phi_k=\phi(\vec{\mu}_k)$, one can quickly find a diagonal form of the generator
$\W^{\free}$. First, we note that the eigenvalue condition gives us 
\be
[\W,\phi_{\pm k}] = \pm 2 \varepsilon_k \phi_{\pm k}.
\ee
Second, these operators satisfy the following algebra,
\be
\{\phi_k,\phi_q\}=2\vec{\mu}_k\cdot\vec{\mu}_q=2\delta_{k,-q},
\ee
which in particular implies also $\phi_k^2=0$. This immediately tells us that, up to a constant
shift, the generator can be given in terms of $\phi_k\phi_{-k}$,
\be
\W^{\free}=\sum_{k=1}^L\varepsilon_k \left(\phi_k\phi_{-k}-2\right) = \sum_{k=1}^{L}
\varepsilon_k \, \phi_{-k}\phi_{k}.
\ee
Here we determined the constant shift so that the eigenvalue of $\W^{\free}$ with the largest
real part is $0$, which ensures the conservation of probabilities.

The diagonalization of the stochastic generator $\W^{\free}$ thus reduces to the
diagonalization of the $2L\times 2L$ block-three-diagonal matrix $M_L$,
\be\label{eq:MLblocktridiagonal}
M_L=
\mkern12mu
\underbrace{
    \mkern-14mu
\begin{bmatrix}
    a_1 & b & & &  & \\
    -b^T  & a & b & & & \\
       & -b^T & a & b & & \\
       & & & \ddots & & \\
       & & & -b^T & a & b \\
       & & &   & -b^T & a_L 
\end{bmatrix}
    \mkern-14mu
}_{L}
\mkern12mu,
\ee
with the $2\times2$ blocks given by
\be
a_1=\frac{\kappa-\delta}{2}\Y{} \qquad
a=\kappa\Y{},\qquad a_L=\frac{\kappa+\delta}{2}\Y{},\qquad
b=\begin{bmatrix}
    -\frac{\kappa-\delta}{2} & 0 \\ -\ii & \frac{\kappa+\delta}{2}
\end{bmatrix}.
\ee

\subsection{Conserved zero modes} \label{sec:ZeroModes}
In this language, conserved zero modes can be understood as eigenvectors of $M_L$ 
corresponding to the eigenvalue $0$. Due to the skew-symmetric structure of the
matrix, they necessarily need to appear in pairs, and we assume the following 
homogeneous ansatz,
\be
\vec{\nu}=\bigoplus_{j=1}^{L} \lambda^{j} \begin{bmatrix}\alpha \\ \beta \end{bmatrix}.
\ee
Requiring $M_L\vec{\nu}=0$, we obtain two linearly independent solutions,
\be
\vec{\nu}_1=\bigoplus_{j=1}^{L} \lambda^{j-1} 
\begin{bmatrix} 1 \\ \ii \lambda\end{bmatrix},\qquad
    \vec{\nu}_2=\bigoplus_{j=1}^{L} 
    \bar{\lambda}^{L-j} \begin{bmatrix} \ii \bar{\lambda} \\ -1\end{bmatrix},
\ee
with the two parameters $\lambda$, $\bar{\lambda}$
\be
\lambda=\frac{1-\sqrt{1-\kappa^2+\delta^2}}{\kappa+\delta},\qquad
\bar{\lambda}=\frac{1-\sqrt{1-\kappa^2+\delta^2}}{\kappa-\delta}
=\left.\lambda\right|_{\delta\leftrightarrow -\delta}.
\ee
We choose the labelling of the eigenvectors $\vec{\mu}_{k}$ so that 
the subscripts $k=1$ and $k=-1$ correspond to $\varepsilon_{\pm 1}=0$, therefore
$\vec{\mu}_{\pm 1}$ should be linear combinations of $\vec{\nu}_{1,2}$, which satisfy
the orthogonality condition~\eqref{eq:orthogonalityLeftRight}. For clarity we take
into account the fact that for all sensible values of $\delta$, $\kappa$, the parameters
$\lambda$, $\bar{\lambda}$ are in magnitude smaller than $1$, and we require the orthogonality
condition to only hold up to corrections of the order $\lambda^{L}$, $\bar{\lambda}^{L}$,
\be
\vec{\mu}_{\pm 1}=\frac{1}{\sqrt{2}}
\bigoplus_{j=1}^{L} \begin{bmatrix}\lambda^{j-1}\pm\bar{\lambda}^{L-(j-1)}\\
    \ii(\lambda^j\pm\bar{\lambda}^{L-j})
\end{bmatrix}.
\ee
At this point it is not clear that these are all the zero eigenvectors.
However, as we will demonstrate later, there are exactly $2L-2$ non-zero
eigenvalues $\pm \varepsilon_{k}$, which means that the degeneracy of the
eigenvalue $0$ is exactly two.

Note that in the main text the zero modes $\Psi$, $\bar{\Psi}$ are required to satisfy
$\Psi^2=\bar{\Psi}^2=1$, and they are not equal to $\phi_{\pm 1}$, but 
they are given by $\vec{\nu}_1$, and $\vec{\nu}_2$. In particular, the expression for
$\Psi^{(FF)}$ from Eq.~\eqref{eq:PsiExpr} is
\be
\Psi=\phi(\vec{\nu}_1)
=\sum_{j=1}^{L}\lambda^{j-1}\left(\A{j}+\ii\lambda\B{j}\right)=
\mu_{j-1}
\Z{1}\Z{2}\ldots \Z{L}\left(\X{j}+\ii \lambda \Y{j}\right),
\ee
while the second zero mode is 
\be
\bar{\Psi}=-\ii\Pi\phi(\vec{\nu}_2)
=\Pi\sum_{j=1}^{L}\bar{\lambda}^{L-j}\left(\bar{\lambda}\A{j} + \ii\B{j}\right)
=\sum_{j=1}^{L}\bar{\lambda}^{L-j}\Z{j+1}\cdots \Z{L}
\left(\X{j}+\ii\bar{\lambda}\Y{j}\right),
\ee
with $\Pi=\prod_{j=1}^{L}\Z{j}$. Here we take advantage of the fact that both $\Pi$,
and $\phi(\vec{\nu}_2)$ commute with $\W^{\free}$, and therefore also their product does.
The two zero modes $\Psi$ and $\bar{\Psi}$ are exponentially localised at left and right edge
respectively. Moreover, $\bar{\Psi}$ is related to $\Psi$ through the left-right reflection,
since this transformation maps $\delta$ to $-\delta$, and
\be
\bar{\lambda}=\left.\lambda\right|_{\delta \leftrightarrow - \delta}.
\ee
%The two zero modes square to one up to exponentially small corrections,
%\be
%\Psi^2=1+\mathcal{O}(\lambda^{2L}),\qquad \bar{\Psi}^2=1+\mathcal{O}(\bar{\lambda}^{2L}).
%\ee

\subsection{Change of basis}
To find the remaining eigenvectors it is convenient to perform a basis transformation.
We first introduce fermionic creation and annihilation operators  $c_{j}$, $c^{\dagger}_j$,
$j=1,\ldots,L-1$,
\be
c_{j}=\frac{1}{2}\left(\B{j}+\ii \A{j+1}\right),\qquad 
c_{j}^{\dagger}=\frac{1}{2}\left(\B{j}-\ii \A{j+1}\right).
\ee
In terms of $c_{j}$, and $c_{j}^{\dagger}$, the generator takes the following form,
\be
\W^{\free}=\ii (\delta-\kappa)\A{1}c_1^{\dagger} + (\delta+\kappa)c_{L-1}^{\dagger}\B{L}
-2\sum_{j=1}^{L-1}c_j^{\dagger}c_j+\sum_{j=1}^{L-2}
\left(
2\kappa c_{j}^{\dagger} c_{j+1}^{\dagger}
+(\delta+\kappa)c_j^{\dagger}c_{j+1}
+(\delta-\kappa)c_j c_{j+1}^{\dagger}
\right).
\ee
Since terms of the form $c_j c_k$, $A_{1} c_j$, and $B_{L} c_j$  are not
present, commutation with a linear combination of $c_j^{\dagger}$ produces
another linear combination of $c_j^{\dagger}$, which means that $L-1$ of the
remaining eigenvectors are just linear combinations of $c_j^{\dagger}$. In other
words, the matrix $M_L$ rewritten in the basis
$\{c_1^{\dagger},c_2^{\dagger},\ldots,c_{L-1}^{\dagger},c_1,c_2,\ldots,c_{L-1},\A{1},\B{L}\}$
has a non-trivial invariant subspace under right action,
\be
\tilde{M}_{L}=
\begin{bmatrix}
    \tilde{a} & \tilde{b} & \tilde{c} \\
    0 & -\tilde{a}^{T} & 0 \\
    0 & -\frac{1}{2}\tilde{c}^{T}& 0
\end{bmatrix},
\ee
where blocks $\tilde{a}$ and $\tilde{b}$ are $L-1\times L-1$, while $\tilde{c}$
is $L-1\times 2$,
\be
\tilde{a}=
\mkern12mu
\underbrace{
    \mkern-14mu
\begin{bmatrix}
    -1 & \frac{\kappa+\delta}{2} & & &  & \\
    \frac{\kappa-\delta}{2} & -1 & \frac{\kappa+\delta}{2} & & & \\
       & \frac{\kappa-\delta}{2} & -1 & \frac{\kappa+\delta}{2} & & \\
       & & & \ddots & & \\
       & & & \frac{\kappa-\delta}{2} & -1 & \frac{\kappa+\delta}{2} \\
       & & &   & \frac{\kappa-\delta}{2} & -1 
\end{bmatrix}
    \mkern-14mu
}_{L-1}
\mkern12mu,\qquad
\tilde{b}=
\mkern12mu
\underbrace{
    \mkern-14mu
\begin{bmatrix}
    0& \kappa & & &  & \\
    -\kappa & 0 & \kappa & & & \\
       & -\kappa & 0 & \kappa & & \\
       & & & \ddots & & \\
       & & & -\kappa & 0 & \kappa \\
       & & &   & -\kappa & 0 
\end{bmatrix}
    \mkern-14mu
}_{L-1}
\mkern12mu,\qquad
\tilde{c}=
\mkern12mu
\underbrace{
    \mkern-14mu
\left.\begin{bmatrix}
    \ \ii(\kappa-\delta) & 0 \\
    0 & 0 \\
    \vdots & \vdots \\
    0 & 0 \\
    0 & \kappa+\delta
\end{bmatrix}
\mkern-4mu\right\}
    \mkern-30mu}_{2}
    \mkern24mu {\scriptstyle L-1}
    \ .
\ee
In this basis the matrix is not skew-symmetric, and therefore the sets of
left and right eigenvectors, $\{\tilde{\vec{\mu}}_k^{L}\}_k$ and
$\{\tilde{\vec{\mu}}_k^{R}\}_k$, are not the same. However, both these sets
can be determined by first diagonalising $\tilde{a}$.
Let $\vec{v}_{k}^{R}$ and $\vec{v}_{k}^{L}$,
be the right and left eigenvectors of $\tilde{a}$ corresponding
to the eigenvalue $-\varepsilon_k$,
\be
\tilde{a}\vec{v}_k^{R}=-\varepsilon_k\vec{v}_k^{R},
\qquad\vec{v}_k^{L}\tilde{a}=-\varepsilon_k\vec{v}_k^{L},\qquad k=2,\ldots L.
\ee
These eigenvalue equations can be straightforwardly solved with the sine transform,
and after some basic manipulations we obtain
\be
\begin{gathered}
\vec{v}_k^{R}=\bigoplus_{j=1}^{L-1}\alpha^j\sin\frac{(k-1)j\pi}{L},\qquad
\vec{v}_k^{L}=\frac{2}{L}\bigoplus_{j=1}^{L-1}\alpha^{-j}\sin\frac{(k-1)j\pi}{L},\qquad
\alpha^2=\frac{\kappa-\delta}{\kappa+\delta}=\frac{1}{\lambda \bar{\lambda}},\\
\varepsilon_{k}=1-(\kappa+\delta)\alpha\cos\frac{(k-1)\pi}{L}.
\end{gathered}
\ee
Note that depending on the magnitude and signs of parameters $\kappa$ and $\delta$, the
eigenvalues $-\varepsilon_k$ can be either complex or real, but the sign of $\varepsilon_k$
was chosen so that $\Re{\varepsilon_k}\ge 0$ for any physically sensible values of parameters.

Using the diagonal basis of $\tilde{a}$ we are now able to immediately obtain \emph{right}
eigenvectors of $\tilde{M}_L$ corresponding to eigenvalues $\varepsilon_{-k}=-\varepsilon_k$,
and \emph{left} eigenvectors of $\tilde{M}_L$ corresponding to eigenvalues $\varepsilon_k$.
Indeed, $\tilde{\vec{\mu}}_{-k}^{R}$ and $\tilde{\vec{\mu}}_k^{L}$ defined as 
\be
\tilde{\vec{\mu}}_{-k}^{R}=\vec{v}^{R}_k\oplus \begin{bmatrix}0\\ \vdots \\ 0\end{bmatrix},
    \qquad
\tilde{\vec{\mu}}_k^{L}=\begin{bmatrix}0\\ \vdots \\ 0 \end{bmatrix}\oplus \vec{v}^{R}_k
    \oplus \begin{bmatrix} 0 \\ 0 \end{bmatrix},
\ee
satisfy the appropriate eigenvalue equations,
\be
\tilde{M}_L\tilde{\vec{\mu}}_{-k}^{R}=-\varepsilon_k \tilde{\vec{\mu}}_{-k}^{R},\qquad
\tilde{\vec{\mu}}_{k}^{L}\tilde{M}_L=\varepsilon_k \tilde{\vec{\mu}}_k^{L}.
\ee
We note that this completely determines the spectrum of $M_L$ (and
$\tilde{M}_L$): two eigenvalues are zero, $\varepsilon_{\pm 1}=0$, while the
rest are given in pairs $\varepsilon_{\pm k}=\pm \varepsilon_{k}$, $k\ge 2$.

To determine the rest of the right eigenvectors we take the following ansatz
\be
\tilde{\vec{\mu}}_k^{R}=\vec{x}_{k}^{R} \oplus \vec{y}^R_k \oplus \vec{z}^R_k,
\qquad
\vec{x}_k^{R},\vec{y}_k^{R}\in\mathbb{C}^{L-1},\
\vec{z}_k^{R}\in\mathbb{C}^2.
\ee
Requiring that this is a right eigenvector of $\tilde{M}_L$ corresponding to
the eigenvalue $\varepsilon_k$, we obtain the following explicit form,
\be
    \vec{y}_k^{R}=\vec{v}_k^{L},\qquad
    \vec{z}_k^{R}=-\frac{1}{2\varepsilon_k}\tilde{c}^T \vec{v}_k^{L},\qquad
    \vec{x}_k^{R}=\left(\tilde{a}-\varepsilon_k\right)^{-1}
    \left(\frac{1}{2\varepsilon_k}\tilde{c}\tilde{c}^{T}-\tilde{b}\right)\vec{v}_k^{L}.
\ee
Note that the matrix $\tilde{a}-\varepsilon_k$ is invertible for any $k$, since
the spectrum of $\tilde{a}$ is $\{-\varepsilon_k\}_{2\le k \le L}$.
The expression for $\vec{z}_k^{R}$ can be immediately evaluated and yields
\be
\vec{z}_k^{R}=-\frac{(\kappa+\delta)\alpha}{2\varepsilon_k}\sin\frac{(k-1)\pi}{L}
\begin{bmatrix}
    \ii\\
    (-1)^k \alpha^{-L}
\end{bmatrix}.
\ee
To find a convenient form of $\vec{x}_k^{R}$, we express it in the basis of the
eigenvectors of $\tilde{a}$ as,
\be
\vec{x}_k^{R}=\sum_{j=2}^L
\frac{1}{\varepsilon_k+\varepsilon_j} 
\left(
\sum_{l=2}^{L}
\left(\vec{v}_j^{L}\cdot(\tilde{b}-\frac{1}{2\varepsilon_k}\tilde{c}\tilde{c}^{T})\cdot
\vec{v}_l^{R}\right) \left(\vec{v}_l^{L}\cdot \vec{v}_k^{L}\right)\right)
\vec{v}_j^{R}=\sum_{j=2}^{L}\frac{1}{\varepsilon_k+\varepsilon_j}
\left(\sum_{l=2}^{L}(f_{j,l}-\frac{1}{2\varepsilon_k}g_{j,l})h_{l,k}\right)
\vec{v}_j^R,
\ee
where we introduced coefficients $f_{j,l}$, $g_{j,l}$ and $h_{j,l}$ to encode the relevant
vector overlaps. With some straightforward manipulations they can be simplified as,
\be
\begin{aligned}
    f_{j,l}&=\vec{v}_j^{L}\cdot \tilde{b} \cdot \vec{v}_l^{R} = 
\delta_{j,l}\kappa \cos\frac{(j-1)\pi}{L}\left(\frac{1}{\alpha}-\alpha\right)
+\frac{1-(-1)^{j+l}}{2}\frac{2\kappa}{L}
\frac{\sin\frac{(j-1)\pi}{L}\sin\frac{(l-1)\pi}{L}}
{\cos\frac{(j-1)\pi}{L}-\cos\frac{(l-1)\pi}{L}}\left(\frac{1}{\alpha}+\alpha\right),\\
    g_{j,l}&=\vec{v}_j^{L}\cdot \tilde{c}\tilde{c}^{T}\cdot\vec{v}_l^{R}=
\frac{2}{L}\sin\frac{(j-1)\pi}{L}\sin\frac{(l-1)\pi}{L}\left(1+(-1)^{j+l}\right),\\
    h_{j,l}&=\vec{v}_j^{L}\cdot \vec{v}_l^{L}=\frac{4}{L^2}
\frac{(-1)^{l+j}\alpha^2(1-\alpha^4)\sin\frac{(l-1)\pi}{L}\sin\frac{(j-1)\pi}{L}}
{\left(1-2\alpha^2\cos\frac{(l+j-2)\pi}{L}+\alpha^4\right)
\left(1-2\alpha^2\cos\frac{(l-j)\pi}{L}\right)}.
\end{aligned}
\ee
The left eigenvectors $\tilde{\vec{\mu}}_{-k}$ can be determined analogously.

\section{Eigenvectors of the edge mode}\label{sec:eigenvectors}
Since the edge mode squares into identity, $\Psi^2=1$, the vectors of the following form
are eigenvectors corresponding to eigenvalues $+1$ and $-1$,
\be \label{eq:genericEigenvector}
\ket{\alpha^{\pm}_{v}} = \left(1\pm\Psi\right)\ket{v}.
\ee
However, for an eigenvector to be used as an initial state in the protocol described in
the main text,  it also has to represent a valid probability distribution,
i.e.\ all it should be normalized and all its component should be non-negative,
\be
\braket{\fl}{\alpha^{\pm}_v}=1,\qquad \braket{\ul{s}}{\alpha^{\pm}_v}\ge 0.
\ee
This requirement immediately constraints us to \emph{positive} eigenvectors $\ket{\alpha^{+}_v}$,
as $\bra{\fl}$ is up to exponentially small corrections a positive left eigenvector
of both the free-fermionic and the interacting edge mode, which in particular implies
\be
\bra{\fl}(1-\Psi)\ket{v}=0
\ee
for an arbitrary $\ket{v}$. To find appropriate positive eigenvectors, we need
to separately treat the two cases.

\subsection{Free-fermionic eigenvectors}
\subsubsection{Basis of eigenvectors}
Before specialising to the case of positive eigenvectors, let us first construct
a basis of eigenvectors of the form~\eqref{eq:genericEigenvector}.

Choosing $\ket{v}$ in~\eqref{eq:genericEigenvector} to be any canonical basis
state $\ket{v}=\ket{s_1 s_2\ldots s_L}$, one obtains $4^L$ different vectors,
which is twice the dimension of the probability space.  We can explicitly show
that this yields at most $2^L$ linearly independent eigenvectors, by using the
fact that $\ket{\alpha^{+}_v}$ is an eigenvector corresponding to the
eigenvalue $1$,
\be \label{eq:genEigv}
(1+\Psi)\ket{v}=(1+\Psi)\Psi\ket{v},
\ee
which in the example of a canonical basis vector $\ket{v}=\ket{0s_2s_3\ldots s_{L}}$
yields
\be
\begin{aligned}
    (1+\Psi)\ket{0s_2\ldots s_L}=(1+\Psi)\frac{1-\lambda}{\sqrt{1-\lambda^{2L}}}
\ket{1s_2\ldots s_L}\mkern-6mu
    +\sum_{k=2}^{L} (1+\Psi)\psi_k\ket{0s_2\ldots s_L},
\end{aligned}
\ee
where we use the short-hand notation
\be
\psi_k=\frac{\lambda^{k-1}}{\sqrt{1-\lambda^{2L}}}\Z{1}\cdots\Z{k-1}
\left(\X{k}+\ii\lambda\Y{k}\right).
\ee
From here it follows that the eigenvectors of the form $(1+\Psi)\ket{1s_2\ldots s_L}$
can be expressed in terms of eigenvectors obtained from basis states $\ket{0s_2\ldots s_L}$,
\be
\ket*{\alpha^{+}_{1 s_2 \ldots s_L}}
\in\Span\Big(\{\ket*{\alpha^{+}_{0 s_2 \ldots s_L}}\} \cup
    \{\ket*{\alpha^{+}_{0 s_2 \ldots 1-s_j\ldots s_L}}\}_{j=2}^L
\Big).
\ee
This immediately implies that there are at most $2^{L/2}$ linearly independent
eigenvectors of the form $\ket{\alpha^{+}_v}$. An analogous statement holds for
$\ket*{\alpha^{-}_v}$ due to the mapping
$\ket*{\alpha^{-}_v}=\prod_{j=1}^L \X{j}\ket*{\alpha^{+}_v}$.

To prove that there are no additional eigenvectors of $\Psi$ we should show that the
set of vectors
\be \label{eq:basisAlphaP}
\{\ket{\alpha^{+}_{0s_2s_3\ldots s_L}} \}_{s_j\in\{0,1\}}
\ee
is linearly independent. We observe that each one of
vectors $\ket{\alpha^{+}_{0 s_2 s_3\ldots s_L}}$ has a nonzero overlap
with precisely one basis vector of the form $\ket{1 s_2^{\prime}\ldots s_L^{\prime}}$,
\be
\braket{1s_2^{\prime} \ldots s_L^{\prime}}{\alpha^{+}_{0s_2 \ldots s_L}}
=\prod_{j=2}^{L}\delta_{s_j^{\prime},s_j^{\phantom{\prime}}}
\mel{1 s_2 \ldots s_L}{\psi_1}{0 s_2 \ldots s_L},
\ee
which directly implies the linear independence of the set~\eqref{eq:basisAlphaP}.

\subsubsection{Positivity requirement}
We start by observing that eigenvectors $\ket{\alpha^{+}_{0\ldots00}}$ and
$\ket{\alpha^{+}_{0\ldots 01}}$ consist of nonnegative components, therefore any
(appropriately normalized) linear combination of these two eigenvectors is a valid probability
distribution. To find the general condition for a positive eigenvector $\ket{\alpha}$ to fulfil
the positivity requirement, we can express it as a linear combination of basis eigenvectors,
\be
\ket{\alpha}=
\sum_{s_2,s_3,\ldots,s_L}
c_{0s_2s_3\ldots s_L} (1+\Psi) \ket{0 s_2\ldots s_L}.
\ee
The straightforward observation is that all the coefficients $c_{0 s_2\ldots s_L}$ must be
non-negative, which follows from
\be
\braket{1s_2 s_3\ldots s_L}{\alpha}=
\frac{1}{\sqrt{1-\lambda^{2L}}} c_{0 s_2 s_3\ldots s_L}.
\ee
The rest of the components are expressed as a sum of $L+1$ contributions,
\be
    \braket{0s_2s_3\ldots s_L}{\alpha}= c_{0 s_2 s_3\ldots s_L}+
    \sum_{j=2}^L \frac{\lambda^{j-1}(-1)^{s_2+\cdots+s_{j-1}}}{\sqrt{1-\lambda^{2L}}}
c_{0 s_2\ldots s_{j-1} (1-s_j) s_{j+1}\ldots s_L},
\ee
and the positivity condition reduces to $2^{L-1}$ inequalities. Note that
two of them are automatically satisfied, since
$\braket{00\ldots0}{\alpha}\ge0$ and $\braket{00\ldots01}{\alpha}\ge 0$
follow from $c_{0 s_2\ldots s_L}\ge 0$.

In general we are unable to be more explicit. However, it is still possible to
find a few more instances of allowed states. For example, the linear
combination
\be
c_{0\ldots010} \ket{\alpha^{+}_{0\ldots 010}}+
c_{0\ldots011} \ket{\alpha^{+}_{0\ldots 011}},
\ee
is positive whenever the ratio between the two coefficients is in the following range,
\be
\frac{(1-\lambda)\lambda^{L-1}}{\sqrt{1-\lambda^{2L}}}
\le \frac{c_{0\ldots 010}}{c_{0\ldots 011}}\le
\frac{\sqrt{1-\lambda^{2L}}}{(1+\lambda)\lambda^{L-1}}.
\ee

In the numerical calculations in the main text we choose 
$\ket{\alpha}$ to be proportional to the simplest basis state
$\ket{\alpha^{+}_{00\ldots 0}}$,  which takes the following explicit form,
\be
\begin{aligned}
    \ket{\alpha} &= \frac{1+\lambda^L-\sqrt{1-\lambda^{2L}}}{2\lambda^L}
    \ket{00\cdots 0}
    +
    \frac{1-\lambda}{1-\lambda^L+\sqrt{1-\lambda^{2L}}}
    \sum_{j=1}^{L}
    \lambda^{j-1} \ket*{\underbrace{0\cdots0}_{j-1}10\cdots0}\\
    &\approx\frac{1}{2} \ket{00\cdots 0}
    +\frac{1-\lambda}{2}\sum_{j=1}^L \lambda^{j-1}
    \ket*{\underbrace{0\cdots0}_{j-1}10\cdots0}.
\end{aligned}
\ee
In the large system-size limit this state has a nice intuitive interpretation: with
probability $\frac{1}{2}$ the initial configuration is empty, and with probabilities
decaying as $\lambda^j$ the initial configuration consists of one particle at position
$j$.

\subsection{Interacting eigenvectors}
As in the free-fermionic case, at the moment we are unable to fully characterize the
set of all positive eigenvectors with non-negative components, but we can provide
a simple family of states that belong to it. In particular, we introduce the state
$\ket{\alpha_k}$ parametrized with an integer $1\le k \le L$ and a probability
parameter $0\le \alpha \le 1$, as 
\be
\ket{\alpha_k} = \frac{1}{2} (1+\Psi)
\left(
\begin{bmatrix}
    \frac{1}{2} \\[0.3em] \frac{1}{2} 
\end{bmatrix}^{\otimes k-1}
\otimes \begin{bmatrix} \alpha\\1-\alpha\end{bmatrix}
    \otimes
\begin{bmatrix}
    \frac{1}{2} \\[0.3em] \frac{1}{2} 
\end{bmatrix}^{\otimes L-k}
\right),
\ee
i.e.\ the vector $\ket{v}$ in~\eqref{eq:genericEigenvector} corresponds to the maximum-entropy
state everywhere, except at position $k$, where the probability distribution is parametrized
by $\alpha$. In the computational basis, it takes the following form, 
\be
    \ket{\alpha_k}=
\sum_{\ul{s}}\frac{1}{2^{L-1}}
\Big(
\alpha + \delta_{s_k,1} (1-2\alpha) 
    +\frac{(1-\lambda^2)(1-2\alpha)}{2}
\sum_{j=1}^{L-1} \lambda^{j+k-2} (-1)^{s_j}
\Big)
\ket{\ul{s}},
\ee
and one can straightforwardly verify that each term is non-negative for any combination of 
parameters $-1<\lambda<1$, and $0\le \alpha \le 1$.

In the numerical simulations in the main text we use
$\ket{\alpha}\equiv\ket{\alpha_1}$, with an intermediate value of $\alpha$
($\alpha=0.25$). 

\section{Stationary states}\label{sec:stationary}
\subsection{Free-fermionic regime}
%\textcolor{purple}{Add a sentence about the stationary state from diagonalization} 
A stationary state~$\ket{s}$ is mapped to $0$ under the stochastic
generator $\W$,
\be \label{eq:stationarity}
\W\ket{s}=0.
\ee
In the case with $\delta=0$, a stationary state can be found in product
form, 
\be
\ket{s}=
\begin{bmatrix}
    \displaystyle
    \frac{1+\kappa}{1+\kappa+\sqrt{1-\kappa^2}}\\[1.5em]
    \displaystyle
    \frac{\sqrt{1-\kappa^2}+\kappa-1}{2\kappa}
\end{bmatrix}^{\otimes L}=
\begin{bmatrix}
    \displaystyle
    \frac{1+\lambda}{2}\\[1em]
    \displaystyle
    \frac{1-\lambda}{2}
\end{bmatrix}^{\otimes L},
\ee
where $\lambda$ is given by the $\delta\to 0$ limit of the expression
in~\eqref{eq:lambdaDef}. This immediately gives the stationary
expectation value of an arbitrary product of $\Z{j}$,
\be
\mel{\fl}{\Z{j_1}\Z{j_2}\cdots \Z{j_k}}{s}=\lambda^k,
\ee
which coincides with the expectation value in the edge-mode eigenvector
(cf.~\eqref{eq:FFrelation}).

We remark that the stationary state is \emph{not} unique, due to the
generator~$\W$ exhibiting a $\mathbb{Z}_2$ symmetry,
\be
[\Pi,\W]=0,\qquad
\Pi=\prod_{j=1}^L \Z{j}.
\ee
There are two linearly independent stationary states and we can choose
the basis of stationary states to consist of (normalised) eigenvectors
of~$\Pi$,
\be
\ket{s_{+}}=\frac{1+\Pi}{1+\lambda^L}\ket{s},\qquad
\ket{s_{-}}=\frac{1-\Pi}{1-\lambda^L}\ket{s},
\ee
so that $\Pi\ket{s_{\pm}}=\pm \ket{s_{\pm}}$. Up to exponentially small
corrections the expectation values of local observables in $\ket{s_{\pm}}$
match the ones in $\ket{s}$, therefore the discussion in the main text
holds regardless of the precise state to which the system relaxes (as long
as $L-k$ is not small). Furthermore, by requiring the stationary expectation
value of $\Pi$ to match the initial value $\mel{\fl}{\Pi}{\alpha}$, one can show
that the stationary state $\ket{s_{\alpha}}$ to which the system relaxes
after starting from $\ket{\alpha}$ matches $\ket{s}$ up to exponentially small corrections,
\be
    \ket{s_{\alpha}}=
    \frac{1}{2}\left(1+\frac{1-\sqrt{1-\lambda^{2L}}}{\lambda^L}\right)\ket{s_{+}}\\
    +
    \frac{1}{2}\left(1-\frac{1-\sqrt{1-\lambda^{2L}}}{\lambda^L}\right)\ket{s_{-}}\\
    =\ket{s} + \mathcal{O}(\lambda^L).
\ee

\subsection{Interacting generator}
In the case of the interacting generator, the stationary state is the maximum-entropy state,
\be
\ket{s}=\frac{1}{2^L} \ket{\fl},\qquad
\W^{\text{(int)}} \ket{s}=0.
\ee
This can be easily understood by noticing that the left/right hopping rates are the same
(therefore the state should be translationally invariant), and the rates for
annihilation/creation of pairs are the same, therefore we expect the probability of finding
a particle at some site the same as not finding it. Alternatively, one can check explicitly
and repeat the discussion above.
The expectation value of the observable $\chi$ [cf.~\eqref{eq:defChi}] is
\be
\mel{\fl}{\chi}{s}=\sum_{j=1}^{L}\lambda^j \mel{\fl}{\Z{j}}{s}=0,
\ee
which follows from $\mel{\fl}{\Z{j}}{\fl}=0$ for any $j$.

\section{Stationary dynamical correlations}\label{sec:correlationME}
In the case of interacting generator the stationary state $\ket{s}=2^{-L}\ket{\fl}$ is 
also a right eigenvector of the edge-mode,
\be \label{eq:rightEigenvector}
\Psi \ket{\fl}=\ket{\fl},
\ee
which enables us to use a similar manipulation to find restrictions on dynamical
correlation functions. We start with the correlation function 
$\mel{\fl}{A e^{tW} B}{s}$ between two observables $A$, and $B$ at times $0$ and $t$,
and take into account~\eqref{eq:rightEigenvector} and~\eqref{eq:flatPsi}, 
\be
\begin{aligned}
  \mel{\fl}{A e^{tW} B}{\fl}&=\mel{\fl}{A e^{t W} B \Psi}{\fl}\\
  &=\mel{\fl}{A e^{tW} \{B,\Psi\}}{\fl}-\mel{\fl}{A e^{t W}\Psi B}{\fl}\\
  &=\mel{\fl}{A e^{tW} \{B,\Psi\}}{\fl}-\mel{\fl}{A\Psi e^{tW} B}{\fl}\\
  &=\mel{\fl}{A e^{tW} \{B,\Psi\}}{\fl}-\mel{\fl}{\{A,\Psi\} e^{tW} B}{\fl}
  +\mel{\fl}{\Psi A e^{tW} B}{\fl}\\
  &=\mel{\fl}{A e^{tW} \{B,\Psi\}}{\fl}-\mel{\fl}{\{A,\Psi\} e^{tW} B}{\fl}
  +\mel{\fl}{A e^{tW} B}{\fl},
\end{aligned}
\ee
which immediately implies the following,
\be\label{eq:relation}
\mel{\fl}{\{A,\Psi\}e^{t W} B}{\fl}=\mel{\fl}{A e^{t W}\{B,\Psi\}}{\fl}.
\ee
We note that this does not necessarily represent a relation between two correlation
functions, as $\{A,\Psi\}$ and $\{B,\Psi\}$ are in general non-diagonal. However,
we can extend the definition of the corresponding diagonal
operator~\eqref{eq:defCorrDiag} for a generic operator $O$ as,
\be
\bra{\fl}\hat{O}_L=\bra{\fl}O,\qquad
\hat{O}_R\ket{\fl}=O\ket{\fl},
\ee
which enables us to rewrite~\eqref{eq:relation} as a relation between two genuine
correlation functions,
\be \label{eq:corrfunRelation}
\mel{\fl}{A e^{t W} \hat{\{B,\Psi\}}_R}{\fl}=
\mel{\fl}{\hat{\{A,\Psi\}}_L e^{t W} B}{\fl}=
\mel{\fl}{B e^{t W} \hat{\{A,\Psi\}}_L}{\fl}.
\ee
The second equality follows from $W^{T}=W$.

Specializing now to the case $A=\Z{j}$ and $B=\Z{k}$ we first note that the
diagonal operators appearing in~\eqref{eq:corrfunRelation} read as,
\be
\hat{\{\Z{j},\Psi\}}_L=2 \Z{j} - 2 \lambda^j\left(\frac{1}{\lambda^2}-1\right)\chi,
\qquad
\hat{\{\Z{k},\Psi\}}_R=2 \Z{k} - 2 \lambda^k\left(\frac{1}{\lambda^2}-1\right)\chi,
\ee
with $\chi$ defined in~\eqref{eq:defChi}. This finally gives us
\be
\lambda^j \mel{\fl}{\Z{k} e^{t W} \chi}{s}=\lambda^k \mel{\fl}{\Z{j} e^{t W} \chi}{s}.
\ee
\end{document}